\documentclass[
  reprint,
  amsmath,amssymb,
  floatfix,superscriptaddress,showkeys
]{revtex4-2}

\usepackage{gensymb}
\usepackage{setspace}
\usepackage{dirtytalk}
\usepackage{natbib}
\usepackage{amsmath}
\usepackage{mathtools}
\usepackage{upgreek}
\usepackage[pdftex]{graphicx}
\usepackage{float}
 \usepackage[left]{lineno}
\usepackage{graphicx}% Include figure files
\usepackage{dcolumn}% Align table columns on decimal point
\usepackage{bm}% bold math

\usepackage{hyperref}% add hypertext capabilities

\begin{document}
%\linenumbers

\preprint{APS/123-QED}

\title{Orbitally resolved single-photon emission \\ from an individual atomic vacancy center in a semiconductor}% Force line breaks with \\
%\thanks{A footnote to the article title}%

\author{Gagandeep Singh}
\affiliation{Division of Physics and Applied Physics, School of Physical and Mathematical Sciences, Nanyang Technological University, Singapore 637371, Singapore}

\author{Xiaodan Lyu}
 \affiliation{Division of Physics and Applied Physics, School of Physical and Mathematical Sciences, Nanyang Technological University, Singapore 637371, Singapore}

\author{Bi Qi Chong}
\affiliation{Division of Physics and Applied Physics, School of Physical and Mathematical Sciences, Nanyang Technological University, Singapore 637371, Singapore}

\author{Ryan Li Yen Tang}
\affiliation{Division of Physics and Applied Physics, School of Physical and Mathematical Sciences, Nanyang Technological University, Singapore 637371, Singapore}

\author{Rejaul SK}
\affiliation{Division of Physics and Applied Physics, School of Physical and Mathematical Sciences, Nanyang Technological University, Singapore 637371, Singapore}

\author{Yande Que}
\affiliation{Division of Physics and Applied Physics, School of Physical and Mathematical Sciences, Nanyang Technological University, Singapore 637371, Singapore}

\author{Ranjith Shivajirao}
\affiliation{Division of Physics and Applied Physics, School of Physical and Mathematical Sciences, Nanyang Technological University, Singapore 637371, Singapore}

\author{Thasneem Aliyar}
\affiliation{Division of Physics and Applied Physics, School of Physical and Mathematical Sciences, Nanyang Technological University, Singapore 637371, Singapore}

\author{Radha Krishnan}
\affiliation{Division of Physics and Applied Physics, School of Physical and Mathematical Sciences, Nanyang Technological University, Singapore 637371, Singapore}
 
\author{Junxiang Jia}
\affiliation{Division of Physics and Applied Physics, School of Physical and Mathematical Sciences, Nanyang Technological University, Singapore 637371, Singapore}

\author{Michael S. Fuhrer}
\affiliation{School of Physics and Astronomy, Monash University, Clayton, Victoria 3800, Australia}

\author{Teck Seng Koh}
\affiliation{Division of Physics and Applied Physics, School of Physical and Mathematical Sciences, Nanyang Technological University, Singapore 637371, Singapore}

\author{Weibo Gao}
\affiliation{Division of Physics and Applied Physics, School of Physical and Mathematical Sciences, Nanyang Technological University, Singapore 637371, Singapore}
 
\author{Bent Weber}
\email{b.weber@ntu.edu.sg}
 \affiliation{Division of Physics and Applied Physics, School of Physical and Mathematical Sciences, Nanyang Technological University, Singapore 637371, Singapore}

%\date{\today}% It is always \today, today,
             %  but any date may be explicitly specified

\begin{abstract}
Atomically confined spins are emerging as active components in quantum optoelectronic devices such as quantum bits and sensors. However, interrogating single spins at atomic length-scales remains a sizeable challenge, limited by diffraction in conventional optics. Here we show that the highly-local excitation provided by injecting energetic charge carriers from the atomically sharp probe of a scanning tunneling microscope can trigger single-photon emission from individual atomic vacancy centers in a layered semiconductor. With an effective spatial resolution of $<$1~nm, we show that the captured light closely mirrors the orbital symmetry of the bound-state wavefunction of the vacancy center while photon correlation measurements confirm single-photon emission, as reflected in clear photon anti-bunching signatures. Our results constitute an important step toward the realization of an electrically addressable single-atom quantum light source and solid-state spin-photon interface, addressed at the atomic-scale.

\end{abstract}


%\keywords{Suggested keywords}%Use showkeys class option if keyword
                              %display desired
\maketitle

%\tableofcontents

\section{Introduction}

Point defects in semiconductors and insulators, capable of confining individual charges and spins, are well-defined atomic-scale quantum systems with applications as optically addressable qubits and single-photon emitters \cite{yin2013optical, aharonovich2016solid}. However, addressing these optically active quantum systems at the individual-level remains a formidable challenge. Conventional optical spectroscopy is fundamentally constrained by the diffraction limit, typically restricting the spatial resolution to laser spot sizes of approximately 300 nm in diameter \cite{schwarz2016electrically}. As a result, measurements are often limited to ensembles of defects leaving the exact number of emitters and their atomic and electronic structure often uncertain, especially when characterizing novel emitter types \cite{tran2016quantum, zhou2018room}. Addressing emission centers at the scale of the defect's Bohr radius ($\approx 1$~nm) would not only allow identification but also deterministic interrogation. Ultimately, this may enable new realms of quantum control and scalable quantum information processing, for instance towards photon-mediated entanglement between pairs of emitters — a cornerstone for scalable quantum networks \cite{knaut2024entanglement, cilibrizzi2023ultra, awschalom2018quantum}.

Different from diffraction-limited conventional optical spectroscopy, the injection of energetic charge carriers from the tip of a scanning tunneling microscope (STM) constitutes a highly local excitation, giving rise to photon emission in scanning tunneling luminescence (STL). STL has been demonstrated for various molecular \cite{zhang2017electrically, grosse2016nanoscale} and atomic-scale systems \cite{zrenner1999spatially, hoffmann2001luminescence, haakanson2002photon} with spectrally and temporally resolved photon emission down to the Angstrom-scale. STL showing single-photon emission has been so far limited to molecular systems \cite{zhang2017electrically, roslawska2018single}. However, applications in quantum communication, computing, and sensing will likely require single-photon sources and interfaces that are embedded within crystalline matrix of a semiconductor or insulator.

Uniquely, the atomic thinness of two-dimensional (2D) materials, such as hexagonal boron nitride (hBN) \cite{tran2016quantum} and the transition metal dichalcogenides (TMDCs) \cite{koperski2015single} enables STL to probe atomic-scale emitters with Angstrom-scale spatial resolution. Semiconducting TMDCs, in particular, exhibit a direct bandgap, large spin-orbit coupling and spin-valley selectivity \cite{yao2008valley}---properties which can be inherited by defect centers \cite{krishnan2023spin, wang2020spin}. This makes TMDCs a promising platform to realize novel spin-photon interfaces \cite{montblanch2023layered}. Quantum emission has been reported for various TMDC species, and has been  attributed to local potentials induced by strain \cite{branny2017deterministic}, atomic defects \cite{he2015single}, or their combination \cite{parto2021defect}. Recently, chalcogen vacancies centers have been identified as quantum emitters \cite{mitterreiter2021role} in photoluminescence experiments, demonstrating spatially localized emission with narrow-band spectrum along with spin-valley selectivity \cite{hotger2023spin} and longer lifetime compared to free excitons in TMDCs \cite{RTsvac}.

\begin{figure*}
\centering
\includegraphics[width=\textwidth]{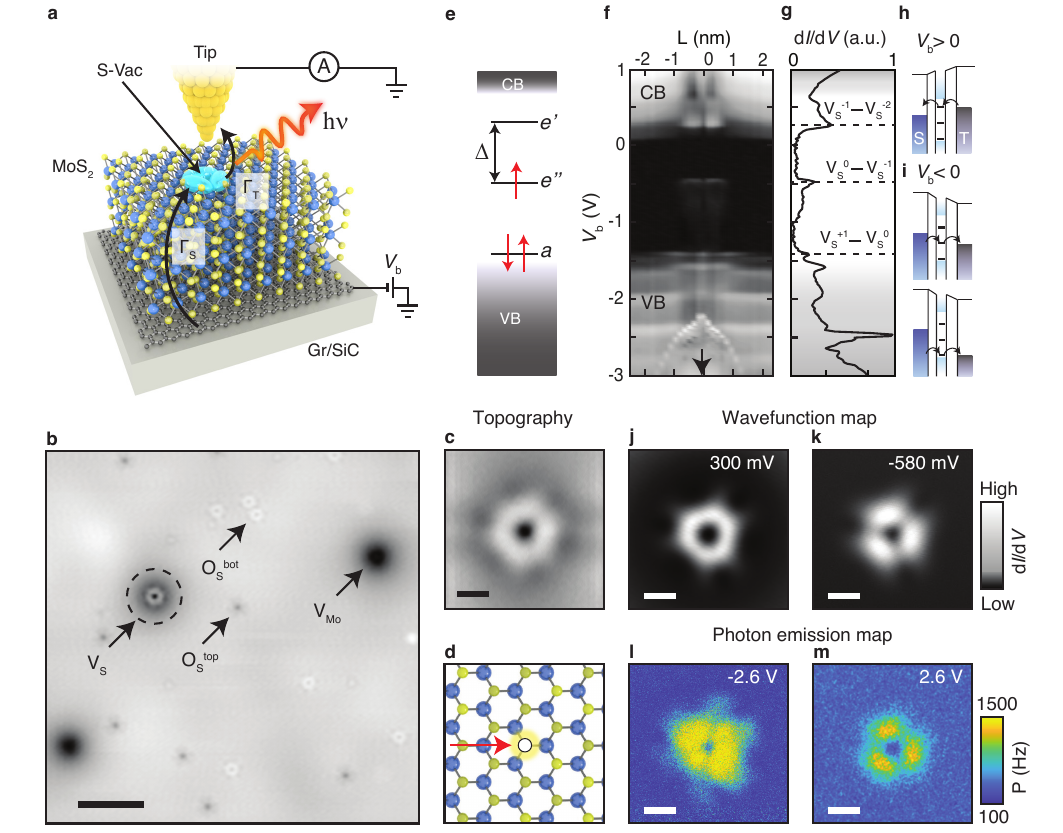}
\caption{\textbf{Orbitally-resolved photon emission from an individual V$_{\rm{S}}$ center in MoS$_2$.} \textbf{a,} Schematic diagram of the experiment showing a single V$_{\rm{S}}$ center in a three layer thick MoS$_2$ crystal on a Gr/SiC substrate, probed by a STM tip. Electrons tunnel from graphene reservoir (biased) to the defect at a rate of $\Gamma_{\rm{S}}$ and from the defect to the tip (grounded) at a rate of $\Gamma_{\rm{T}}$. \textbf{b,} STM topographic image of the MoS$_2$ surface, recorded at $V_{\rm{b}} = 0.7$~V, showing a variety of defect species including a sulfur vacancy (V$_{\rm{S}}$), molybdenum vacancy (V$_{\rm{Mo}}$) and oxygen passivated sulfur vacancies in the top (O$\rm{_{S}^{top}}$) and sub-surface (O$\rm{_{S}^{bot}}$). \textbf{c-d,} Close-up topographic image of a V$_{\rm{S}}$ recorded at $V_{\rm{b}} = 0.7$~V alongside a corresponding lattice schematic. \textbf{e-f,} Energy-level diagram and spectroscopic measurement of the V$_{\rm{S}}$ center's in-gap electronic structure. \textbf{g,} Point spectrum taken atop a V$_{\rm{S}}$ reflecting three in-gap states: \textbf{a, e'} and  \textbf{e''}. \textbf{h-i,} Energy-level diagram showing resonant tunneling via the unoccupied \textbf{e'}, the occupied \textbf{e''} and \textbf{a} defect level, respectively. Here, S and T denote the substrate and tip reservoirs, respectively. \textbf{j-k,} dI/dV wavefunction maps of the \textbf{e'} and \textbf{e''} states. \textbf{l-m,} Photon emission maps recorded at $V_{\rm{b}} = -2.6$~V and $2.6$~V, respectively, at a constant current of 10~nA, closely mirroring the defect wavefunctions in (j, k). Scale bar: 5~nm in (b) and 6~\AA~in (c, j-m).}
\label{fig1}
\end{figure*}

To date, several attempts have been made to capture STL from 2D semiconductors \cite{roman2020tunneling, pommier2019scanning,   pechou2020plasmonic, lopez2023tip, krane2016electronic, schuler2020electrically, phark2025roadmap}, including excitonic emission \cite{lopez2023tip, pommier2019scanning, roman2020tunneling, pechou2020plasmonic}. Recent experiments have achieved STL that is spatially localized to individual S-vacancies in WS$_2$ \cite{schuler2020electrically}, confirming them as promising candidates for an atomic-scale light source. Yet, while providing spectral detail on the energetics of the emission, no single-photon emission has been demonstrated.

Different from previous studies \cite{schuler2020electrically} here we focus explicitly on the temporal statistics of emitted photons from individual defect centers, thus allowing us to present direct evidence of atomically resolved single-photon emission (SPE) for the first time in a semiconductor host. Exploring the sulfur vacancy (V$\rm{_S}$) in MoS$_2$, as a model system, we show that SPE in STL is triggered by single-charge tunneling through discrete V$\rm{_S}$-induced quantum states within the bandgap, dictated by Coulomb blockade. Single-photon emission manifests twofold, (1) in a saturation of the photon emission rate as a function of pump power (injected tunneling current) at low energy ($eV_b\approx h\nu$) coinciding with (2) a clear photon anti-bunching signature in photon-correlation measurements using Hanbury-Brown-Twiss interferometry. The quantized nature of photon emission is further confirmed at larger energy ($eV_b>2h \nu$) by the observation of photon bunching with $g^{(2)}(0)\approx 21$. Our work provides a powerful demonstration of resolving light-matter interaction down to the atomic-scale and up to GHz frequencies. As a method applicable to the broader class of 2D semiconductors and insulators, this technique constitutes a significant step towards the realization of on-demand electrically driven single-atom quantum light sources, and a method to identify novel classes of emitters electro-optically, at atomic-length scales.

\section{Results}

A schematic of the experiment is given in Fig. \ref{fig1}a showing an atomically sharp Au tip probing a V$_{\rm{S}}$ center in a few-layer MoS$_2$ crystal on monolayer graphene (see Methods). Electronic in-gap states induced by the vacancy center form a double barrier junction with the STM tip above (grounded), via the vacuum, and with graphene counter electrode below (biased), via the van-der-Waals gap. In this biasing configuration, electrons tunnel from filled states of the defect to the tip at negative bias and vice-versa, and charge tunneling is dominated by charging of the microscopic junction capacitances leading to Coulomb blockade \cite{aliyar2024symmetry}. In response to single-charge tunneling, photons of energy $h\nu$ are emitted, and detected in the far-field by an in-situ optical lens placed in proximity to the STM junction (refer SI for details on the experimental setup).

\begin{figure*}
\centering
\includegraphics[width=\textwidth]{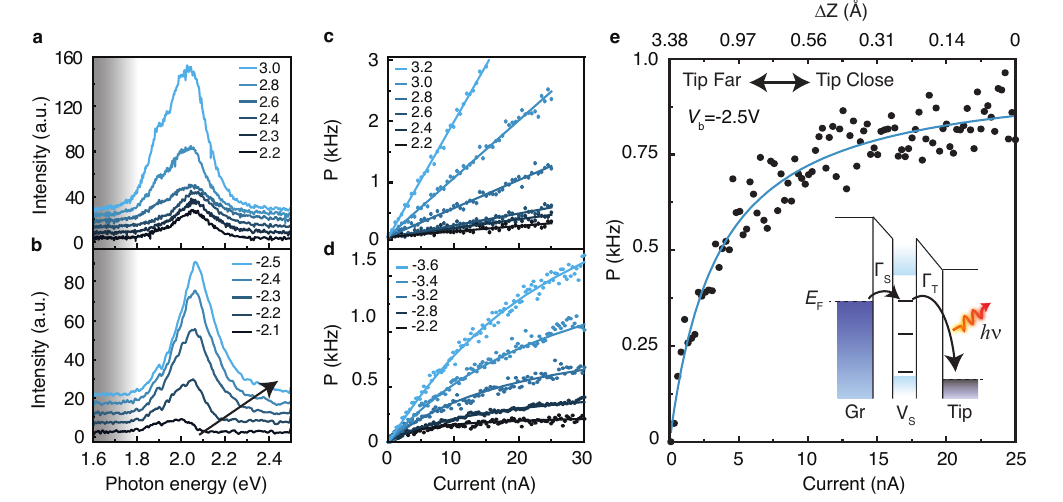}
\caption{\textbf{Atomically-resolved optical emission spectra and excitation dependence.} \textbf{a-b,} Optical spectra at a constant excitation current of 10~nA at positive polarity (a) and negative (b) bias. \textbf{c-d,} Photon emission rate as a function of tunneling current for different biases at (c) positive bias polarity exhibiting linear dependence (solid traces are guides to the eye) and at (d) at negative polarity (solid traces are fits to Eq. 16 in SI).  The saturation observed in the photon emission rate arises when photon emission is dominated by tunneling through discrete in-gap states of the defect. \textbf{e,} A fit to the two-state rate equation model (Eq. 2), we extract the charge tunnel rates $\Gamma_{\rm{S}}$ and $\Gamma_{\rm{T}}$. Inset: Energy-level diagram of the emission mechanism (see main text for a description). Grey shading in (a, b) represents lower transmission of the setup. Data in (d, e) are from separate measurements using different STM tips.}
\label{fig2} 
\end{figure*}

An STM topographic image of the MoS$_2$ surface is shown in Fig.~\ref{fig1}b, revealing a variety of atomic defect species. V$_{\rm{S}}$ vacancy centers (dashed black circle) can be identified from their characteristic ring-like structure \cite{aliyar2024symmetry, xiang2024charge}. A missing sulfur atom leaves three Mo dangling bonds behind in the lattice (Fig. \ref{fig1}d). These give rise to a manifold of three in-gap states (Fig. \ref{fig1}e), split into $\textbf{a}$ (occupied) and $\textbf{e}$ (unoccupied) states by the crystal field \cite{tan2020stability, naik2018substrate}. Previous work \cite{aliyar2024symmetry} from our group confirmed the orbital contributions by density functional theory (DFT) and showed that the \textbf{a} state is dominated by Mo d$_{\rm{xz}}$ and d$_{\rm{yz}}$ orbitals whereas the \textbf{e} manifold primarily consists of Mo d$_{\rm{xy}}$ and d$_{\rm{x^2-y^2}}$ orbitals. In the pristine lattice, sulfur is present in a S$^{2-}$ ionization state. Removal of a sulfur atom therefore frees up two electrons that pair up to form a charge-neutral spin singlet in the lower energy $\textbf{a}$-state near the valence band. As previously shown \cite{aliyar2024symmetry}, the high $\textit{n}$-doping present in our samples allows for the occupation of the  $\textbf{e}$ state manifold with an unpaired spin-1/2, negatively charging the vacancy defect (V$_{\rm{S}}^{-1}$). A charge-state dependent Jahn-Teller (JT) distortion, lifting the $\textbf{e}$ state degeneracy and breaking the three-fold symmetry of the wavefunction, is predominantly observed in monolayer MoS$_2$ crystals \cite{aliyar2024symmetry, xiang2024charge}. In the multilayer crystals investigated here,  most V$_{\rm{S}}$ defects (approx. 10 out of 12) exhibit three-fold orbital symmetry.

\begin{figure*}
\centering
\includegraphics[width=\textwidth]{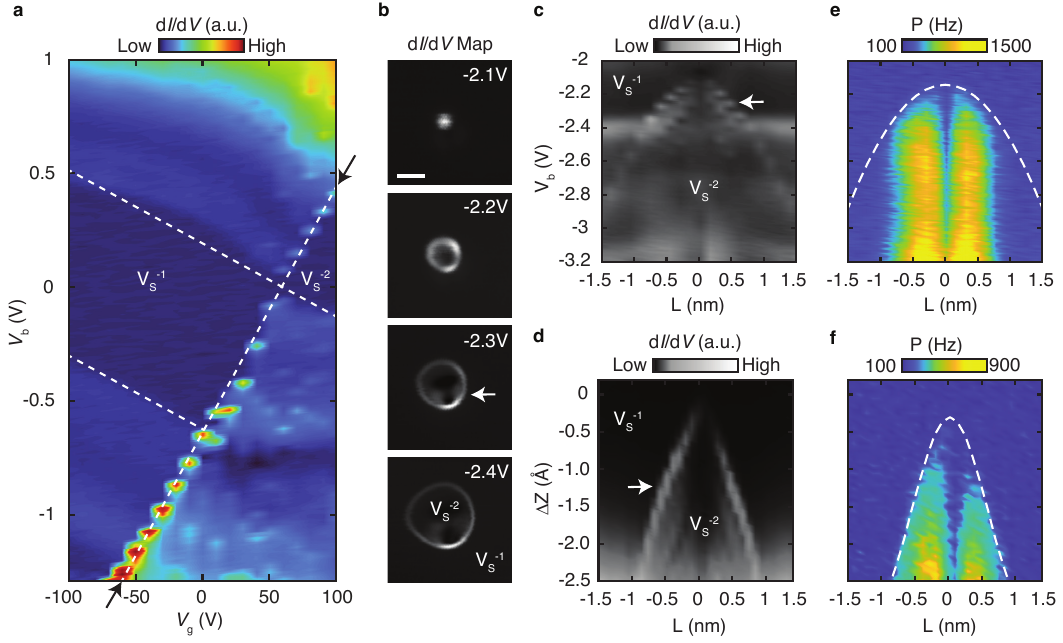}
\caption{\textbf{Charge-state dependent photon emission.} \textbf{a,} Charge stability diagram (Coulomb diamond) of an individual V$\rm{_S}$ center recorded in a back-gated device (refer to SI for device structure). White dashed lines indicates the alignment of the V$\rm{_S}$ electrochemical potential with the tip and graphene Fermi level allowing conduction via the in-gap states (refer SI for details). Current is Coulomb-blockaded within the dashed white boundaries at low bias. Arrow indicates the alignment of the V$\rm{_S}$ electrochemical potential $\mu_{\rm{V_{S} ^{-1}\leftrightarrow V_{S} ^{-2}}}$ with the sample Fermi-level allowing a change in the equilibrium charge state (charging peak). The data was obtained at a constant tip height with a lock-in modulation of 10~mV. The discretizations of the charging peak is due to a large $V_{\rm{g}}$ step size of 10~V. \textbf{b,} Mapping of the charging peak at constant tip height reveals a ring-like structure centered around the V$\rm{_S}$ whose size increases with applied bias. \textbf{c-f,} Simultaneously recorded d$I$/d$V$ and photon count rates at a V$\rm{_S}$ center, as a function of tip-defect lateral ($L$) and vertical ($\Delta z$) distance, at fixed tip height in (c, e) and fixed bias $V_{\rm{b}}=-2.5$V in (d, f). Scale bar in (b) is 1~nm.}
\label{fig3}
\end{figure*}

The V$_{\rm{S}}$ in-gap electronic structure is confirmed in Fig.~\ref{fig1}f,g from differential conductance (d$I$/d$V$) measurements across an individual vacancy center showing three distinct in-gap peaks. The strong suppression near $E_{\rm{F}}$ reflects the presence of the Coulomb blockade gap. As shown in \cite{aliyar2024symmetry}, this confirms single-charge tunneling via  V$\rm{_S} ^{-1}$ $\leftrightarrow$ V$\rm{_S} ^{-2}$ charge transitions at positive, and V$\rm{_S} ^{0}$ $\leftrightarrow$ V$\rm{_S} ^{-1}$ as well as V$\rm{_S} ^{+1}$ $\leftrightarrow$ V$\rm{_S} ^{0}$ charge transitions at negative bias (Fig. \ref{fig1}h,i). The inverted parabola, visible in Fig. \ref{fig1}f, results from tip-induced band bending inducing a change of the equilibrium charge state of the V$_{\rm{S}}$ defect from V$\rm{_S} ^{-1}$ to V$\rm{_S} ^{-2}$, similar to previous observations in GaAs \cite{lee2011tunable}, graphene \cite{brar2011gate}, and other TMDCs \cite{schuler2019large, aliyar2024symmetry, di2022defects}. Further detail on this signature and its role in photon emission is given in Fig. \ref{fig3}.

Differential conductance maps of the local density of states $\rho = \sum_{i} |\psi (x)|^2 \delta(E_i-E)$ are shown in Fig.~\ref{fig1}j,k, reflecting the defect wavefunctions for different charge states. Corresponding photon emission maps (Fig. \ref{fig1}l,m) confirm that luminescence is confined to within $<1$~nm of the vacancy center, closely mirroring the defect wavefunction's orbital symmetry. The photon emission map at positive (negative) bias polarity resemble the wavefunction map of \textbf{e"} (\textbf{e’}) state, indicating the role in-gap states in the emission mechanism. First principle calculations have shown that the \textbf{e}-manifold primarily consists of contribution from Mo d$_{\rm{xy}}$ and d$_{\rm{x^2-y^2}}$ orbitals \cite{aliyar2024symmetry}. The distinct spatial textures observed in the STL maps at positive and negative bias reflect the contribution of these orbitals and their differential coupling to tip’s orbitals. Photon emission maps recorded at different bias (refer SI Fig. 7) show that emission is localized to defect site up to $V\rm{_b}>-4~V$ (for negative bias polarity) and $V\rm{_b}> 2.8~V$ (for positive bias polarity), beyond which emission becomes delocalized across the MoS$_2$ surface. This occurs when charge tunneling no longer proceeds via the confined in-gap states, but via the 2D conduction or valence bands \cite{schuler2020electrically}.

%Further confirmation of the defect-localized emission is shown in SI, confirming that the photon emission at both positive (<~xx~eV) and negative bias (<~yy~eV) remains confined to the defect site at low biases and beyond which it becomes delocalized.Figure \ref{fig2}k-n shows schematics for the inelastic transitions responsible for photon emission in the different bias regimes, respectively.

In STL, photon emission can arises from one of two mechanisms, either excitonic or plasmonic in nature \cite{kuhnke2017atomic}. In excitonic emission, electron-hole pairs created by charge injection at the junction can subsequently radiatively recombine to emit photons \cite{grosse2017submolecular}. In plasmonic emission, inelastically tunneling electrons themselves provide energy to excite plasmon modes at the tunnel junction which subsequently decay radiatively \cite{kuhnke2017atomic}. Excitonic emission usually exhibits narrow spectra of only few meV in width (depending on excitonic lifetime) corresponding to the optical bandgap. Plasmonic emission is generally broadband (a few eV), depends on the dielectric function of tip or substrate \cite{fojtik2003photon, kuhnke2017atomic, berndt1993electromagnetic}, and its intensity varies linearly with excitation rate $P(I)$.

The photon emission spectra and $P(I)$ curves in Fig.~\ref{fig2} confirm a plasmonic emission mechanism. At both bias polarities, emission is broadband with photon energies spanning several hundred meV. The spectra are attenuated by a high-energy cut-off (indicated by arrow in Fig. \ref{fig2}b), $h\nu \leq eV_{\rm{b}}$, at which the photon energy matches that provided by the applied bias \cite{lambe1976light, martin2020unveiling}. An apparent low energy cut-off ($h \nu <1.8$~eV) is imposed by the optical transmission efficiency of the setup (see SI for details). Beyond biases $eV_{\rm{b}}>2.3$~eV, photon emission is further limited by a larger dissipation of plasmons generated due to an increase in the frequency-dependent imaginary part of the dielectric function of the Au tip \cite{hoffmann2001influence}. In general, we find that the spectral detail thus depends on the exact geometrical shape of the tip \cite{aizpurua2000role} (refer SI Fig. 8b).
 
As expected from plasmonic emission mechanism, $P(I)$ is linear for a positive bias polarity and at large  negative bias $V\rm{_b}\leq-4V$. However, in stark contrast with this expectation at low negative biases, we observe a clear saturation of the photon count rate. Such saturation is usually seen as a signature of photon anti-bunching \cite{merino2015exciton}, as commonly observed in conventional optical spectroscopy of quantum emitters \cite{tran2016quantum, kurtsiefer2000stable}.

We can reconcile the coexistence of plasmonic emission with a single-photon character by the observation that the defect center's charging dynamics are dictated by single-charge tunneling in Coulomb blockade. That is, charges tunnel one by one from the graphene to the tip via the defect-bound quantum states at low-bias $V\rm{_b}\ >-4V$. The individual tunneling electrons subsequently excite plasmonic modes at the tip that decay to emit single photons. The energy of these photons is given by the energy difference ($h\nu\leq\Delta E$) between the initial and final states of the inelastically tunneling electrons at the tip-side of the junction (inset Fig. \ref{fig2}e).

We can estimate the photon emission rate by modelling the single-charge transport through the defect (see SI section S8). Solid traces in Fig. 2d show fits to the data. We find that the charge transport and hence photon emission can be simplified to the familiar two-state rate equation model with an effective tunnel rate ($\Gamma_{\rm{eff}}$). The photon emission rate $P$ can be written as

\begin{equation}
    P = \eta_{\rm{C}} \eta_{\rm{P}} \Gamma_{\rm{eff}} = \eta_{\rm{C}} \eta_{\rm{P}} \frac{\Gamma_{\rm{S}} \Gamma_{\rm{T}}}{\Gamma_{\rm{S}}+\Gamma_{\rm{T}}}. 
    \label{fitphotoncurr}
\end{equation}

Here, $\eta_{\rm{C}}$ and $ \eta_{\rm{P}}$ are the collection efficiency of the setup and the efficiency of plasmonic emission, respectively. Further considering that $\Gamma_{\rm{T}}$ depends exponentially on tip height $\Delta z$, the rate of photon emission becomes

\begin{equation}
    P =
     \eta_{\rm{C}} \eta_{\rm{P}} \frac{1} {\Gamma_{\rm{S}}^{-1}+\Gamma_{\rm{0}}^{-1} \rm{exp}{(2\kappa\Delta z)}}.
    \label{fitphotoncurr}
\end{equation}

Simultaneously recording tip height ($\Delta z$) with tunneling current and photon emission rate, we can fit to the observed saturation in photon emission rate as shown by the solid blue trace in Fig. \ref{fig2}e, from which we extract effective charge tunnel time, $\uptau_{\rm{eff}} = 1/\Gamma_{\rm{eff}} = 0.14 \pm 0.00042$~ns at the saturation. The maximum photon count rate achievable at any fixed bias is governed by the electron-to-photon conversion efficiency of the gold tip ($\eta_{\rm{P}}$), which is intrinsically sensitive to the tip geometry. As shown in the SI Fig. 8a, we observe the saturation across multiple vacancy centers and STM tips. While the absolute saturation count rates vary---consistent with differences in $\eta_{\rm{P}}$---the bias range over which saturation occurs is consistent across data sets.

\begin{figure}
\centering
\includegraphics{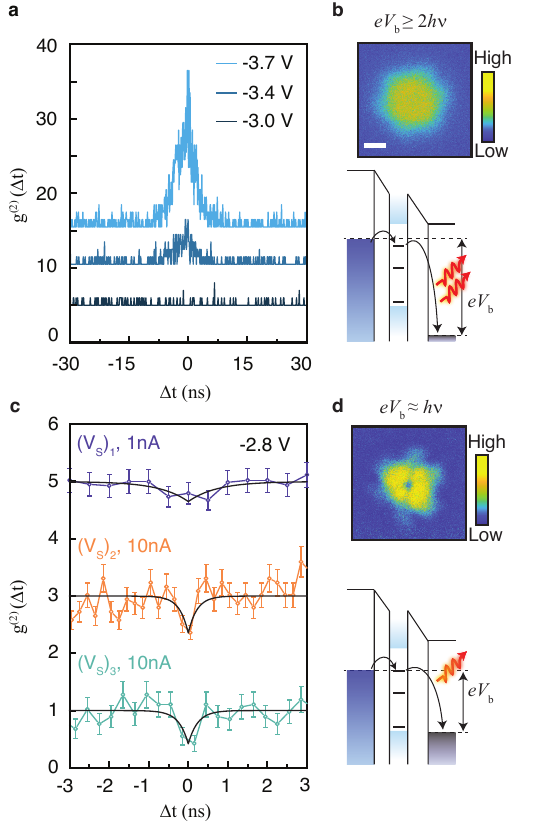}
\caption{\textbf{Single-photon and photon-pair emission.} \textbf{a, c,} Second-order correlation function as a function of time-delay measured at high bias (a) $eV_{\rm{b}}> 2h\nu$ and low bias (c) $V_{\rm{b}}=-2.8~$V. Here, $\rm{(V_{S})_{1}, (V_{S})_{2}, (V_{S})_{3} }$ refer to three different vacancy centers. The parameters extracted from fits are $\uptau$~=~0.71, 0.23, 0.24 ns and g(2)(0) = 0.65, 0.36, 0.42, respectively for the three vacancy centers. \textbf{b, d} STL map (normalized photon count rate) and energy level schematic illustrating the emission mechanism corresponding to the bias regime. A vertical offset of 5 in \textbf{a} and 2 in \textbf{c} has been added for clarity. Inelastic tunneling from other in-gap states may also give rise to photon emission, however these transitions are not shown as the resultant photons may not be within the detectable energy range. Error bars in (c) are the first standard deviation.}
\label{fig4}
\end{figure}

To further confirm single-charge tunneling in Coulomb blockade, we have recorded the defect center's charge stability plot (Coulomb diamonds) in gate-voltage dependent spectroscopy. Fig. \ref{fig3}a, showing the differential conductance over a gate voltage ($V_{\rm{g}}$) range of $-$100 V to 100 V, clearly demonstrate gate-control of the defect's electrochemical potential. The pronounced peak in the charge stability diagram (arrow in Fig. \ref{fig3}a) indicates the alignment of the vacancy center's $\mu_{\rm{V_{S} ^{-1}\leftrightarrow V_{S} ^{-2}}}$ electrochemical potential with the graphene Fermi level, corresponding to the tip-induced charging feature seen in Fig. \ref{fig1}f,g. This conclusion is further confirmed in Fig. \ref{fig3}b-d where we investigate the spatial and bias dependence of this feature (charging rings). Photon emission is strongly enhanced immediately upon a change of the equilibrium charge state from $\rm{V_{S} ^{-1}}$ to $\rm{V_{S} ^{-2}}$ likely due to a strong dependence of the charge tunneling rates on the defect center's charge state, allowing us to record optical spectra (Fig. \ref{fig2}b) only upon defect charging.

To test for photon correlations and ultimately confirm the single-photon character in emission, we have measured the second-order photon correlation function $g^{(2)}(t)$ (Fig. \ref{fig4}) using a Hanbury-Brown and Twiss interferometer. At large bias $eV_{\rm{b}} \geq 2h\nu$, photon bunching indicates photon-pair emission as evident from positive correlations at zero time delay. We understand this as direct evidence of plasmonic emission, similar to observations \cite{leon2019photon} in STL of metal-insulator-metal (MIM) tunnel junctions. Here, the large bias allows for the emission of multiple photons, with energy $h\nu_1 + h\nu_2=eV_{\rm{b}}$, while conjectured that spontaneous parametric down-conversion of plasmon polaritons, enabled by the optical non-linearity of the nanometer tunnel junction, could potentially be responsible for the observed photon bunching \cite{leon2019photon}. The high bunching ratio $\rm{g^{(2)}(0)=21}$  ($eV_{\rm{b}}=-3.7$~eV) observed indicates super-bunching, different from what would be expected for a simple chaotic light source ($g^{(2)}(0)\leq2$). This may further be a consequence of the interplay of photon-pair production and Coulomb blockade, generating single-photon pairs separated in the time domain by charge tunneling events.

%We note that the peak value $g^{(2)}(0)$ is dependent on the time resolution of the setup ($\delta$~=~1.4~ns) \cite{ann2015correction}. As a guide to the eye, we plot an exponential function: $g^{(2)}(\rm{t})=[1+ \textit{g}^{(2)}(0) exp(-\rm{t/t}_0)]$ with time constant $\uptau$ equal to the instrument response time ($\delta_{\rm{setup}}$) of our setup (dashed black trace). The close match of instrument response and experimental data reveals that the measurement is limited by the time resolution of the setup and $g^{(2)}(0)$ maybe even larger if measured at increased temporal resolution.

Building up photon emission statistics from a single atomic vacancy at low bias ($V_{\rm{b}} > -3~V$) is extremely challenging \cite{schuler2020electrically} due to the overall low photon emission count rate and the comparatively low collection efficiency of the UHV-STM based setup (see SI). Despite these challenges, in this work we demonstrate clear photon antibunching signatures, as shown in Fig. \ref{fig4}c. The data was recorded after optimization of our setup including the replacement of the thermal shield's IR filter to include a larger spectral range, optimization of the optical alignment, and the implementation of an atom-tracking module compensating for thermal drift. The data shows separate measurements of three different V$_{\rm{S}}$ centers in a MoS$_2$ bilayer, with clear evidence of photon anti-bunching as evident from a dip at zero time-delay down to g$^{(2)}$(0) $\approx$ 0.36. The absence of single-photon emission at positive bias is likely due to higher tunneling rates resulting in photon emission dynamics far exceeding the temporal resolution of our $g^{(2)}$ setup. From fits to the $g^{(2)}$ curves, we extract effective lifetimes of 0.23 ns and 0.24 ns, respectively, in agreement with that extracted from fitting the photon saturation (Fig. \ref{fig2}e) as well as single-charge tunneling times in few-layer TMDCs \cite{bobzien2025layer}. Decreasing the setpoint to 1~nA increases the effective charge lifetime (refer SI section S8) to 0.71~ns at a reduced signal-to-noise ratio, ultimately confirming that single-charge tunneling in the Coulomb blockade regime triggers the photon emission.

Our results reveal that the observed luminescence is largely dominated by plasmonic emission mechanism yet exhibits signatures of single-photon emission, indicating that radiative dynamics is dominated by Coulomb blockade effects associated with charge tunneling through discrete defect states. In contrast to spectrally resolved dipole transitions providing details of fine structure of atomic or molecular orbitals \cite{germanis2025waveguide, siampour2023observation}, the Angstrom-scale spatial resolution offered by STL provides a direct real-space visualization of the orbital symmetries involved. Future work will consider introducing an atomically thin insulator as decoupling layer between the emitter and the metallic substrate to supress the plasmonic contribution. This configuration may ultimately allow to access the radiative transition between two defect-induced states, similar to transitions observed in NV centers in diamond and hBN emitters \cite{siampour2020ultrabright, stern2022room, gruber1997scanning}, which can uncover the local excitonic structure, orbital symmetries and transition dipole characteristics of the defect-based emitters in 2D materials.

\section{Conclusion} 

In conclusion, we have demonstrated orbitally resolved single-photon emission from an individual vacancy center in a layered semiconductor, captured for the first time with Angstrom-scale spatial resolution via local electrical excitation. Single-photon emission originates from inelastic single-charge tunneling through the discrete in-gap quantum states of a vacancy center in MoS$_2$. Single-photon emission is evident from a saturation of photon count rate upon excitation current coinciding with the clear signatures of photon anti-bunching in photon-correlation measurements. Our results are a major step towards an on-demand electrically-driven single-atom photon source for novel emitter types in 2D crystals, and potential spin-photon interface towards applications in quantum communication, computing, and sensing.
\newline
\section{Methods} 

\textbf{Sample Fabrication.} Epitaxial graphene substrates were obtained by flash annealing 6H-SiC(0001) substrates in an ultra-high vacuum (UHV) preparation chamber with a base pressure of $4 \times 10^{-10}$~mbar. Few-layered crystals of MoS$_2$ were exfoliated from the bulk onto polydimethylsiloxane stamps and dry transferred onto the epitaxial graphene. Microscopic gold markers were deposited through a shadow mask and used to guide the STM tip towards the MoS$_2$ crystal. Prior to STM measurement, the samples were annealed at 250\celsius~ for four hours.

\textbf{Scanning tunneling spectroscopy.} Low-temperature scanning tunneling microscopy and spectroscopy (STM/STS) measurements were performed in an Omicron STM at $T\simeq 4.5$~K under UHV conditions ($\approx 5 \times 10^{-11}$~mbar). In all STM measurements, we used an electrochemically etched gold tip calibrated against the Au(111) Shockley surface state. Spectroscopy measurements were performed using standard lock-in techniques with a modulation amplitude of $V_{\rm{mod}}$= 20~mV and a modulation of frequency of 732 Hz unless otherwise specified. Differential conductance maps were taken in constant height mode.

\section{Supporting Information} 
Ex-situ sample characterization, Estimation of tunnel rates, Details of gate-tunable differential conductance spectroscopy, Tip-induced defect charging, STL setup, STL characterization on Au (111), Extended data on STL from a sulfur vacancy, Modeling of the photon emission dynamics.

\section{Acknowledgment} 
This research is supported by the National Research Foundation (NRF) Singapore, under the Competitive Research Programme Towards ``On-Chip Topological Quantum Devices" (NRF-CRP21-2018-0001) with further support from the Singapore Ministry of Education (MOE) Academic Research Fund Tier 3 grant (MOE-MOET32023-0003) ``Geometrical Quantum Materials", and the Air Force Office of Scientific Research under award number FA2386-24-1-4064.

%\section{Author Contributions} 

%GS and JJ fabricated the samples. GS and RS performed the scanning tunnelling spectroscopy experiments. GS, TA made the setup and performed STL on Au(111). GS performed the STL experiments on MoS$_2$. GS and XL did the photon correlation measurements. BW and GS analyzed the data. BW conceived and coordinated the project. GS and BW wrote the manuscript with input from all authors.

\section{Competing Interests}

The authors declare no competing interests.
%\bibliographystyle{achemso}
%\bibliography{reference}

%apsrev4-2.bst 2019-01-14 (MD) hand-edited version of apsrev4-1.bst
%Control: key (0)
%Control: author (8) initials jnrlst
%Control: editor formatted (1) identically to author
%Control: production of article title (0) allowed
%Control: page (0) single
%Control: year (1) truncated
%Control: production of eprint (0) enabled
%

\end{document}